\documentclass[10pt,letterpaper]{article}
\usepackage{opex3}

\usepackage[dvips]{epsfig}
\usepackage{subfigure}

\usepackage{amsfonts}
\usepackage{amsmath}
\usepackage{amssymb}
\usepackage{pslatex}
\usepackage{color}
\usepackage{hhline}
\usepackage{threeparttable}
\usepackage{supertabular}
\usepackage{multirow}
\usepackage{array}
\usepackage{tabularx}
\usepackage{rotating}

\usepackage[mathscr]{eucal} 

\newcolumntype{Y}{>{\centering\arraybackslash}X}

\renewcommand{\arraystretch}{1.5} 
\begin{document}

\title{Design of plasmonic photonic crystal resonant cavities for polarization sensitive infrared photodetectors}


 \author{Jessie Rosenberg$^{1,*}$, Rajeev V. Shenoi$^2$, Sanjay Krishna$^2$, and Oskar Painter$^1$}
 \address{$^1$Thomas J. Watson, Sr.,\ Laboratory of Applied Physics, California Institute of Technology, Pasadena, CA 91125, USA\\
$^2$Center for High Technology Materials (ECE Dept), University of New Mexico, Albuquerque, NM 87106}
\email{jessier@caltech.edu}




\begin{abstract}
We design a polarization-sensitive resonator for use in mid-infrared photodetectors, utilizing a photonic crystal cavity and a single or double-metal plasmonic waveguide to achieve enhanced detector efficiency due to superior optical confinement within the active region. As the cavity is highly frequency and polarization-sensitive, this resonator structure could be used in chip-based infrared spectrometers and cameras that can distinguish among different materials and temperatures to a high degree of precision.
\end{abstract}






\section{Introduction}
\label{sec:intro}


Optical sensors in the mid-infrared wavelength range are extremely important in a wide variety of areas, such as night vision, missile guidance, and biological spectroscopy \cite{Rogalski:2009}. Currently, the best mid-infrared detectors are based on mercury-cadmium-telleuride (MCT). MCT detectors are very efficient, but large-area focal-plane arrays are difficult and expensive to grow due to difficulties with the epitaxial growth of mercury-based compounds \cite{Phillips:2002, Sidorov:1998}. More recently, other detector materials have become more common, but they have various limitations in the required direction of incoming light, such as in quantum well detectors, or in detector efficiency, as in quantum dot or dots-in-a-well (DWELL) detectors \cite{Krishna:PhysD2005}. With the use of a resonant cavity, it becomes possible to increase the detector efficiency many times over by greatly extending the interaction length between the incoming light and the active material. Instead of incoming light making only one pass through the active region, in a resonant detector the light can make hundreds of passes.

The presence of a resonator can also make each pixel frequency and polarization specific \cite{Painter:2002, Loncar:2002, PainterVuckovic:1999}, allowing for a hyperspectral and hyperpolarization sensor without the need for any external grating or prism. There are many applications for frequency and polarization-sensitive detectors. A hyperspectral detector array could function as a spectrometer on a chip, filtering incoming signals through the use of hundreds of highly sensitive detector pixels. A hyperpolarization detector could be used in a camera to provide an additional layer of information which can be combined with frequency and intensity data to better distinguish between different objects in an image.

The resonator system we investigate here is composed of a photonic crystal cavity for in-plane confinement, and a plasmonic waveguide \cite{Raether:1988, Prade:1991}, composed of either a single or a double-layer of metal, for the vertical confinement. This resonator design, combining the benefits of a plasmonic waveguide and a photonic crystal cavity, has a number of advantages. The plasmonic waveguide serves multiple purposes: it serves as a superior top contact (or in the double-metal case, top and bottom contact) for the detector device providing enhanced extraction efficiency; it provides strong vertical confinement (nearly total confinement, for the double-metal structure) of the resonator mode within the active region; and it increases the index contrast in the photonic crystal, enhancing the in-plane confinement of the resonator mode and enabling strong confinement even with a very shallow photonic crystal etch extending only through the top metal layer \cite{Bahriz:2007, RosenbergAPL:2009}. The photonic crystal patterning also serves a dual purpose: it provides in-plane confinement to the resonator mode, serves as a grating coupler to couple normal-incidence light into the in-plane direction of the detector, and provides a mechanism for freely adjusting the polarization response of the detector pixel.

In the past, many promising schemes have been proposed and/or demonstrated illustrating various aspects of these concepts: optical resonators to provide spectral \cite{Homeyer:2009} or spectral and polarization filtering \cite{Laux:2008, Yang:2008, Hu:2008}, enhanced confinement of light to increase material absorption \cite{Homeyer:2009, Yang:2008, Hu:2008, White:2009}, and metallic gratings to enable strong confinement without the necessity for deep etching \cite{Laux:2008, Hu:2008, White:2009}. Previously, we have demonstrated plasmonic photonic crystal designs with the maximum field intensity at the top metal interface to allow for thinner devices and increase field overlap with the active region versus metallic Fabry Perot-based structures, in both deep-etched \cite{Shenoi:2007} and shallow-etched \cite{RosenbergAPL:2009} single-metal-layer implementations. Here we detail the design of the demonstrated shallow-etch single-metal resonators, and expand those design principles to propose a highly-efficient shallow-etch double-metal cavity design for hyperspectral and hyperpolarization, strongly enhanced mid-infrared detection. Both single-metal and double-metal designs are detector material agnostic, are easily incorporated into current FPA processing techniques, and do not involve the damage or removal of any detector active region material, providing significantly increased flexibility and functionality with a minimal increase in complexity.

\section{Photonic Crystal Design}
\label{sec:photonic_crystal}

A significant obstacle to using resonant cavities to enhance detector absorption and provide spectral and polarization sensitivity is achieving sufficient input coupling from free-space light. Commonly such resonators, with their high confinement, have only very poor phase-matching to a normal incidence free-space beam such as that which we would ideally like to detect for imaging applications. However, with suitable design and optimization of the plasmonic photonic crystal structure, it becomes possible to achieve significant free-space coupling, and indeed, even move towards achieving critical coupling (as will be discussed in Section~\ref{sec:critical_coupling}).

We used group theory to design a frequency and polarization sensitive photonic crystal structure suitable for coupling efficiently to normal incidence light. The simplest polarization-sensitive resonator design would be a one-dimensional grating. However, it is beneficial to choose a fully-connected photonic crystal design in order to take full advantage of the increased current extraction efficiency from the plasmonic metal layer serving as the top contact of the detector device, as well as allowing for continuous variation between polarization-sensitive and polarization-insensitive devices. Therefore, we analyze a square-lattice structure here, and describe how stretching the lattice in one direction can split the degenerate modes of the structure and create a strong polarization sensitivity for use in imaging applications.

\begin{figure}[h]
\begin{center}
\includegraphics[width=\columnwidth]{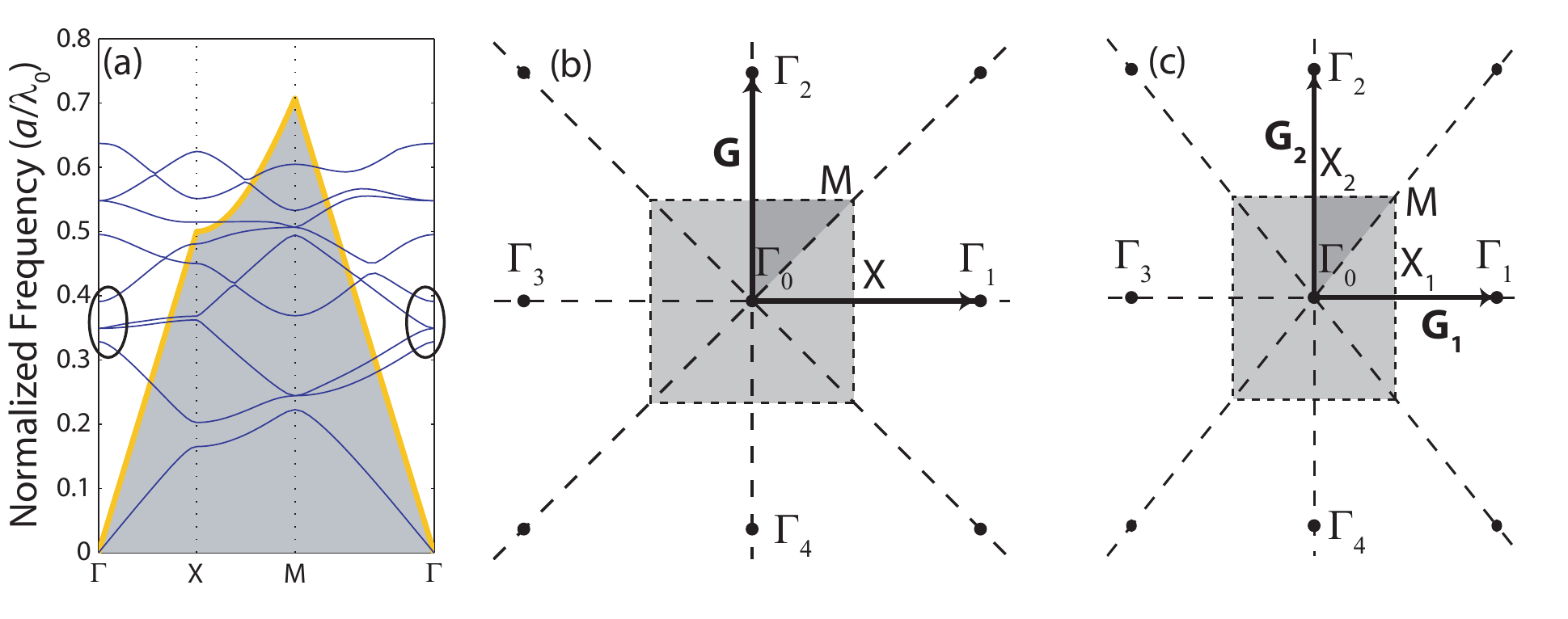}
\caption{In-plane guided mode TM-like bandstructure plot of a square-lattice photonic crystal ($n_{\textrm{eff}} = 3.24$) with $r/a = 0.32$. The light line is shown in yellow, and the modes of interest are circled.} \label{fig:matlab_bandstructure}
\end{center}
\end{figure}

The bandstructure of a square lattice photonic crystal is shown in Fig. \ref{fig:matlab_bandstructure}(a), calculated using plane wave expansion. The ratio of circular hole radius $r$ to lattice spacing $a$ used was $r/a=0.32$, and the index of the material was taken to be the effective index of the double metal plasmon waveguide, $n_{\textrm{eff}}=3.24$. Details of the plasmon effective index calculation are discussed in Section~\ref{sec:plasmon2}. There are no band gaps for this structure, however there are several flat-band regions. The group velocity of these band-edge modes is close to zero, therefore the light travels very slowly and is effectively confined within the patterned region. Band-edge modes are ideal for applications such as detectors, as the mode volume is large, allowing more of the active region to be contained within the resonator. There are a number of flat-band regions within the bandstructure in Fig. \ref{fig:matlab_bandstructure}(a), but we are interested in the modes at the $\Gamma$-point. The $\Gamma$-point corresponds to normal-incidence modulo a reciprocal lattice vector, so the $\Gamma$-point modes are capable of coupling normal-incidence light into the in-plane direction of the detector. In addition, the $\Gamma$-point modes are above the light line, and therefore leak into the air, enabling them to couple more easily to an input free space beam. We investigated the four lowest-order $\Gamma$-point modes, circled in Fig. \ref{fig:matlab_bandstructure}(a), using group theory \cite{Tinkham:2003, Painter:2003}.

The point group symmetry of the square photonic crystal lattice, with reciprocal lattice shown in Fig. \ref{fig:matlab_bandstructure}(b), is C$_{4v}$. The in-plane field of the unperturbed waveguide is given by $\mathbf{E}_{\mathbf{k}_{\perp}}(\mathbf{r}_{\perp}) = \hat{z} e^{- \imath (\mathbf{k}_{\perp}) \cdot \mathbf{r}_{\perp}}$, with $\mathbf{k}_{\perp}$ and $\mathbf{r}_{\perp}$ representing the in-plane wavenumber and spatial position, respectively. When the structure is patterned, coupling will occur between waveguide modes with similar unperturbed frequencies, and propagation constants that differ by a reciprocal lattice vector $\mathbf{G}$.

There is one $\Gamma$-point within the first Brillouin zone (IBZ), at $\{(0,0)k_\Gamma\}$, with $k_\Gamma = 2\pi/a$. Since we are interested in modes with nonzero k-vectors in the in-plane direction, we will consider the nearest $\Gamma$-points in the surrounding Brillouin zones, at $(\{\pm(1,0)k_\Gamma, \pm(0,1)k_\Gamma\})$. These points are labeled in Fig. \ref{fig:matlab_bandstructure}(b). The group of the wave vector, the symmetry group of a plane wave modulo $\mathbf{G}$, is $C_{4v}$ at the $\Gamma$-point. The character table of $C_{4v}$ is shown in Table \ref{tab:character_table}.

\begin{table}
\begin{center}
\caption{Point Group character tables for the square and rectangular lattice.}
\label{tab:character_table}
\begin{tabular}{c|ccccc||c|cccc}
\hline
$C_{4v}$ & $E$ & $C_2$ & $2C_4$ & $2\sigma_v$ & $2\sigma_d$ & $C_{2v}$ & $E$ & $C_2$ & $\sigma_x$ & $\sigma_y$ \\
\hline
$A_1$ & 1 & 1 & 1 & 1 & 1 & $A_1$ & 1 & 1 & 1 & 1 \\
$A_2$ & 1 & 1 & 1 &- 1 & -1 & $A_2$ & 1 & 1 & -1 & -1 \\
$B_1$ & 1 & 1 & -1 & 1 & -1 & $B_1$ & 1 & -1 & -1 & 1 \\
$B_2$ & 1 & 1 & -1 & -1 & 1 & $B_2$ & 1 & -1 & 1 & -1 \\
$E$ & 2 & -1 & 0 & 0 & 0 & & & & & \\
\hline
\end{tabular}
\renewcommand{\arraystretch}{1.0}
\end{center}
\end{table}

\begin{figure}[h]
\begin{center}
\includegraphics[width=0.6\columnwidth]{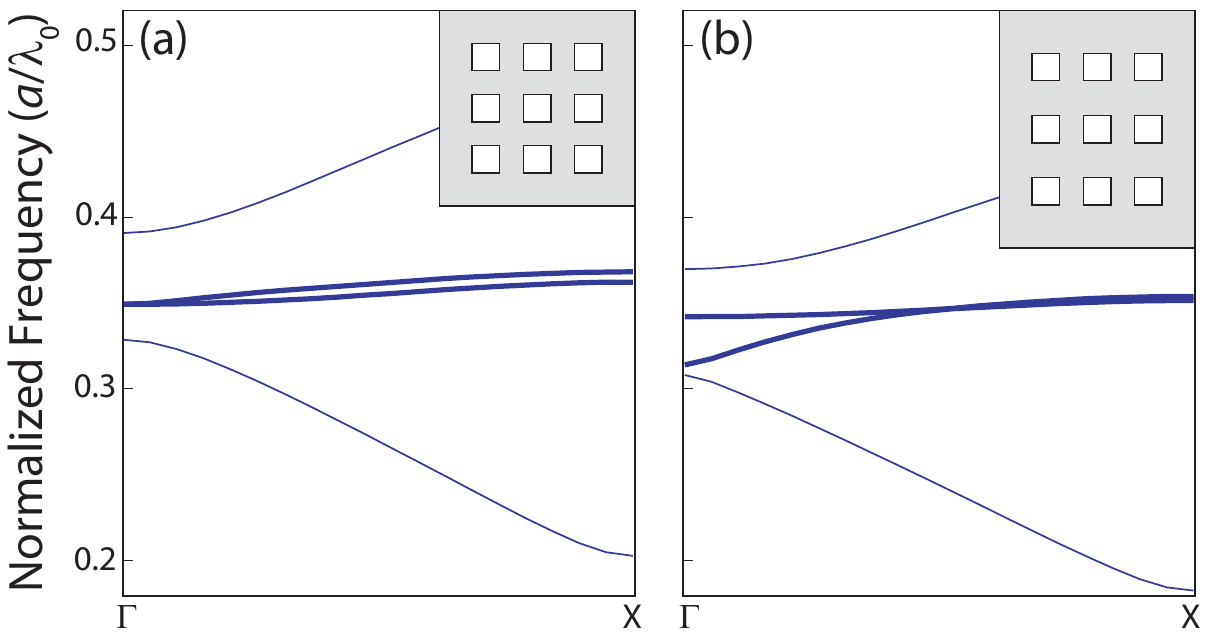}
\caption{2D bandstructure plots near the gamma point of (a) a square lattice and (b) a rectangular lattice, stretched by 10\%. The dipole-like modes are shown in bold.} \label{fig:matlab_bandstructure_stretched}
\end{center}
\end{figure}

The star of $\mathbf{k}$ ($\star\mathbf{k}$) at the $\Gamma$-point is the set of independent $\Gamma$-points within the region. In this case, $\star\mathbf{k}$ is given, not uniquely, by $\{\mathbf{k}_{\Gamma_1}\}$. This will be our seed vector. We find the symmetry basis for the modes at that satellite point by applying the symmetry operations of the group of the wave vector to the seed vector. In this case, the basis is $(\mathbf{E}_{\mathbf{G}_1},\mathbf{E}_{-\mathbf{G}_1}, \mathbf{E}_{\mathbf{G}_2}, \mathbf{E}_{-\mathbf{G}_2})$. Projecting this symmetry basis onto the irreducible representation (IRREP) spaces of $C_{4v}$, we find the modes:
\begin{equation}
\begin{array}{lcl}
\mathbf{E}_{A_1} & = & \hat{z}(\cos(\mathbf{k}_{\mathbf{G}_1} \cdot \mathbf{r}) + \cos(\mathbf{k}_{\mathbf{G}_2} \cdot \mathbf{r})), \\
\mathbf{E}_{B_1} & = & \hat{z}(\cos(\mathbf{k}_{\mathbf{G}_1} \cdot \mathbf{r}) - \cos(\mathbf{k}_{\mathbf{G}_2} \cdot \mathbf{r})), \\
\mathbf{E}_{E,1} & = & \hat{z}(\sin(\mathbf{k}_{\mathbf{G}_2} \cdot \mathbf{r})), \\
\mathbf{E}_{E,2} & = & \hat{z}(\sin(\mathbf{k}_{\mathbf{G}_1} \cdot \mathbf{r})),
\label{eq:group_theory_profile}
\end{array}
\end{equation}
where $A_1$, $B_1$, and $E$ are IRREP spaces of $C_{4v}$ (see Table \ref{tab:character_table}), and $\mathbf{r}$ has its origin at the center of the air hole. Considering that modes with more electric field concentrated in areas with high dielectric constant tend to have lower frequency than those with electric field concentrated in low dielectric regions \cite{ref:Joannopoulos}, we can order the modes by frequency. $E$ is a two dimensional IRREP, so generates two degenerate modes. We associate this pair of modes $\{\mathbf{E}_{E,1}, \mathbf{E}_{E,2}\}$ with the second and third frequency bands, which is in agreement with the bandstructure in Fig. \ref{fig:matlab_bandstructure}(a). These degenerate modes, with dipole-like symmetry and the spatial pattern given in Eqs.~\ref{eq:group_theory_profile}, radiate with a far-field pattern which is uniform: in the case of a finite structure, a Gaussian-like far-field without anti-nodes.

In order to achieve polarization sensitivity, we need to split these two degenerate dipole-like modes. To do this, we stretch the photonic crystal lattice (not the photonic crystal holes) in one direction, giving the reciprocal lattice shown in Fig. \ref{fig:matlab_bandstructure}(c). The effect of stretching the lattice on the four lowest-order $\Gamma$-point modes is shown in Fig. \ref{fig:matlab_bandstructure_stretched}. The symmetry group of this perturbation is $C_{2v}$; the character table for $C_{2v}$ is shown in Table \ref{tab:character_table}. Using the compatibility relations between $C_{4v}$ and $C_{2v}$, we find the new set of modes:
\begin{equation}
\begin{array}{lcl}
\mathbf{E}_{A_1,1} & = & \hat{z}(\cos(\mathbf{k}_{\mathbf{G}_1} \cdot \mathbf{r}) + \cos(\mathbf{k}_{\mathbf{G}_2} \cdot \mathbf{r})), \\
\mathbf{E}_{A_1,2} & = & \hat{z}(\cos(\mathbf{k}_{\mathbf{G}_1} \cdot \mathbf{r}) - \cos(\mathbf{k}_{\mathbf{G}_2} \cdot \mathbf{r})), \\
\mathbf{E}_{B_1} & = & \hat{z}(\sin(\mathbf{k}_{\mathbf{G}_2} \cdot \mathbf{r})), \\
\mathbf{E}_{B_2} & = & \hat{z}(\sin(\mathbf{k}_{\mathbf{G}_1} \cdot \mathbf{r})).
\end{array}
\end{equation}
These modes are plotted in Fig.~\ref{fig:FDTD_group_theory_compare}. The $C_{4v}$ two-dimensional representation $E$ decomposes into $B_1 \oplus B_2$ under $C_{2v}$, therefore the dipole-like modes are no longer degenerate. This is in agreement with what we see in the bandstructure of the stretched lattice, Fig. \ref{fig:matlab_bandstructure_stretched}(b).


\begin{figure}[h]
\begin{center}
\includegraphics[width=\columnwidth]{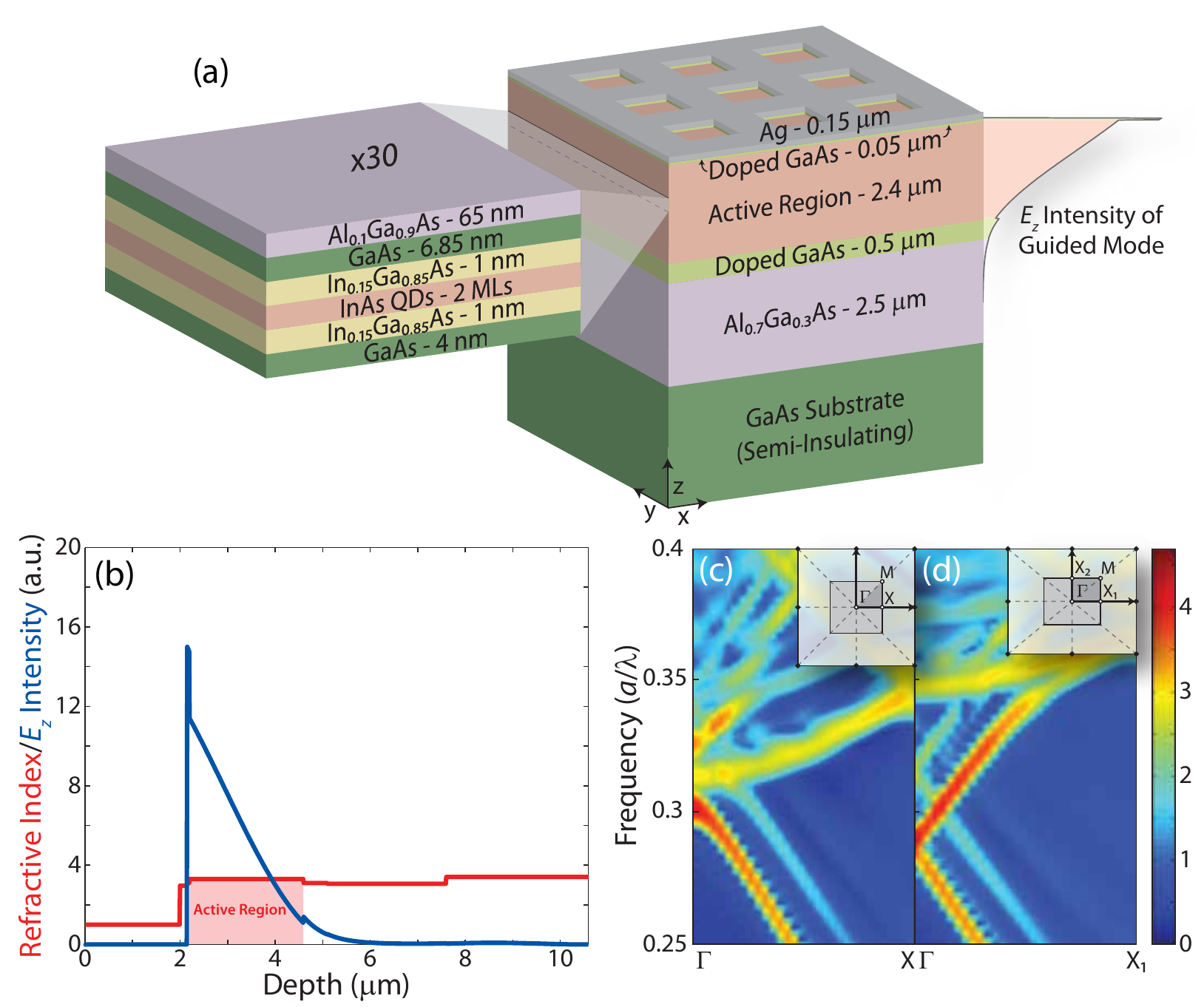}
\caption{(a) A crossectional image of several lattice constants of the single-metal DWELL detector design. (b) $E_z$ intensity profile of the fundamental plasmon waveguide mode (blue) and the real part of the refractive index of the layers (red), with the detector active region highlighted. (c) FDTD bandstructure for the unstretched single-metal photonic crystal structure shown in (a) in the region between the $\Gamma$ and $X$ points. (d) FDTD bandstructure between the $\Gamma$ and $X_1$ points for a single-metal photonic crystal structure stretched and compressed by 10\% in the $x$ and $y$ directions, respectively.} \label{fig:single_metal_crossection}
\end{center}
\end{figure}

\section{Single-Metal Plasmon Resonator Design}
\label{sec:plasmon}

In addition to the in-plane confinement provided by the photonic crystal pattern discussed in Section~\ref{sec:photonic_crystal}, it is necessary to confine the light in the vertical direction as well. We begin with a single-metal design suitable for straightforward fabrication (as experimentally demonstrated in \cite{RosenbergAPL:2009}), and then expand the discussion to consider a double-metal design that mimics the top and bottom contact layers in detector focal plane arrays for easy integration. As we operate in the mid-infrared frequency range, we are far from the plasmon resonance frequency of metals, typically in the ultraviolet; operating in this regime avoids the very high metal losses that occur at frequencies closer to the plasmon frequency, and allows the mode to extend farther into the active region of the detector.

The single-metal resonant DWELL detector structure we study is shown in Fig.~\ref{fig:single_metal_crossection}(a), along with the 1D $E_z$ intensity profile of the fundamental plasmon waveguide mode in Fig.~\ref{fig:single_metal_crossection}(b) (not including the effects of the photonic crystal holes). The vertical confinement factor of this mode within the active region of the detector is $\eta = 91$\%, in strong contrast to the generally much lower confinement factor of purely dielectric waveguides. This extremely high confinement, even for a waveguide utilizing only a single layer of metal, immediately showcases the benefits of choosing a plasmon-based photonic crystal design.

The TM square lattice single-metal plasmonic photonic crystal bandstructure for the region near the $\Gamma$-point is shown in Fig.~\ref{fig:single_metal_crossection}(c), calculated using finite difference time domain (FDTD) methods with metal material properties from Ref.~\cite{Johnson:1972}. The simulated structure has a lattice constant $a$ of $2.939~\mu$m, a hole width vs. lattice constant ratio ($W/a \equiv \bar{W}$) of 0.567, and a metal thickness of $t_m = 150$~nm, and shows $\Gamma$-point modes which are in good agreement with the group theory predictions in the previous section. Figure~\ref{fig:single_metal_crossection}(d) shows the same bandstructure region for a lattice stretched and compressed by 10\% along the $\hat{x}$- and $\hat{y}$- axes respectively, with the two dipole modes showing a significant frequency splitting, also as predicted. These two 3D FDTD simulations match up well with the 2D plane-wave expansion bandstructure predictions shown in Fig.~\ref{fig:matlab_bandstructure_stretched}, with the addition of visible higher-order vertical modes from the single-metal plasmon waveguide. The double-metal waveguide structure can be designed such that these higher-order vertical modes are eliminated.

\begin{figure}[h]
\begin{center}
\includegraphics[width=\columnwidth]{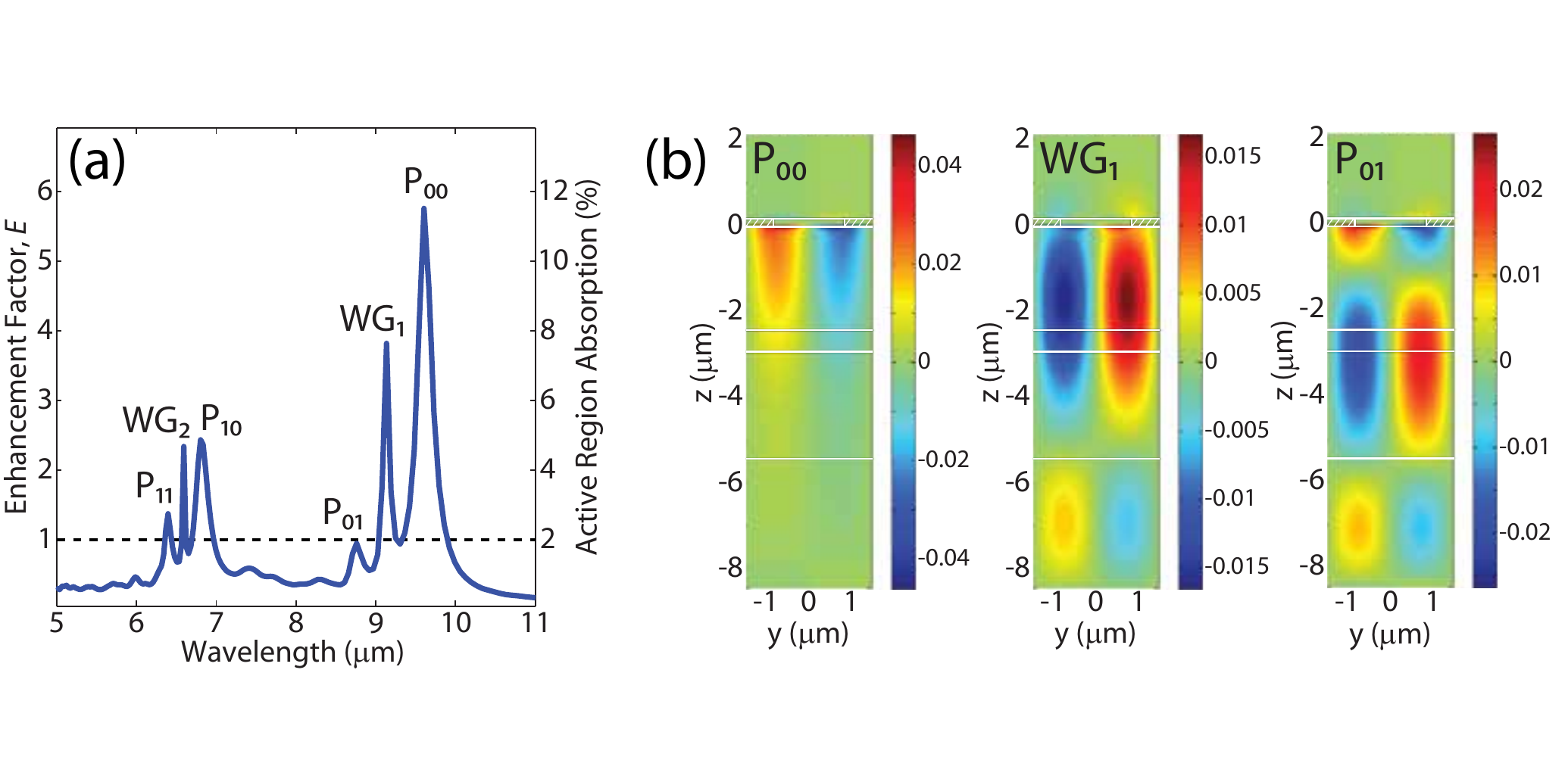}
\caption{(a) FDTD simulated enhancement factor and active region absorption vs. wavelength, based on a 2\% single-pass absorption, using a structure with lattice constant $a = 2.939~\mu$m, $\bar{W} = 0.567$, and metal thickness $t_m = 150$~nm. (b) $E_z$ mode profiles in the y-z plane at the hole edge for the three longer-wavelength peaks in (a), for one lattice constant. The three shorter-wavelength peaks have similar vertical field profiles, but are higher-order in the x-y plane.} \label{fig:enhancement_factor}
\end{center}
\end{figure}

The overall quality factor of the two degenerate dipole modes in the unstretched-lattice case (for resonator parameters given in the caption to Fig.~\ref{fig:enhancement_factor}) is calculated to be $Q_{\textrm{per}} = 48$ for a perfectly periodic structure, not including the in-plane quality factor, $Q_{xy}$, which can be increased indefinitely by adding more lattice constants to the resonator structure. By simulating the structure as excited by an incoming, normal incidence plane wave, we can measure the percentage of the incoming light absorbed within the active region. The simulated DWELL active region absorption corresponding to this mode is $A_t = 11.5$\%, with material parameters specified so as to reproduce the approximate DWELL material single-pass absorption, $A_{\textrm{DWELL}} = 2$\%. This corresponds to an expected responsivity enhancement factor of $E \equiv A_t/A_{\textrm{DWELL}} = 5.75$ at the resonant wavelength, versus a sample with no top patterned plasmonic metal layer at the same wavelength (Fig.~\ref{fig:enhancement_factor}). Note that this enhancement factor is not normalized to the area of the holes in the plasmonic metal, as is typically done in the case of 'extraordinary' transmission through thin metal layers \cite{Barnes:2003}; the absorption values are compared over the same physical region of detector material.

The active region absorption, $A_t$, and enhancement factor, $E$, are plotted versus frequency in Fig.~\ref{fig:enhancement_factor}(a), showing the fundamental plasmon mode at 9.6~$\mu$m and a series of higher-order modes at shorter wavelengths. The mode $P_{xy}$ corresponds to the $x^{\textrm{th}}$-order in-plane and $y^{\textrm{th}}$-order vertical plasmon waveguide mode. Thus $P_{00}$ corresponds to the fundamental plasmon mode as discussed above. $WG_1$ and $WG_2$ are the first and second-order TM waveguide modes of the structure, and Fig.~\ref{fig:enhancement_factor}(b) shows that they have a lower overlap with the metal surface than the plasmon modes. Thus their $Q_per$ is higher than the plasmon modes for the perfectly periodic structure simulated, as can be seen in Fig.~\ref{fig:enhancement_factor}(a), but their in-plane $Q$-factor, $Q_{xy}$, will be very small due to their minimal interaction with the photonic crystal grating. As the overall absorption enhancement factor is lower for these modes even in the infinite-structure limit, we conclude that the response of this resonator structure will be dominated by the surface plasmon-guided modes.


\section{Double-Metal Plasmon Resonator Design}
\label{sec:plasmon2}

Though we have shown that a single-metal plasmonic device can have high active-region confinement and reasonable $Q$-factors, we can move to a double-metal design to increase both of these quantities even further. The double-metal structure brackets the active region with a thin layer of plasmonic metal on either side, and the photonic crystal holes are etched only into the top metal layer, as before. All of the advantages of the single-metal device are preserved, while the substrate loss can be essentially eliminated and the detector active region vertical confinement can approach 100\%. These are achieved at the price of higher plasmonic metal loss, but in the mid-infrared region, this loss is not prohibitive.

\begin{figure}[h]
\begin{center}
\includegraphics[width=0.75 \columnwidth]{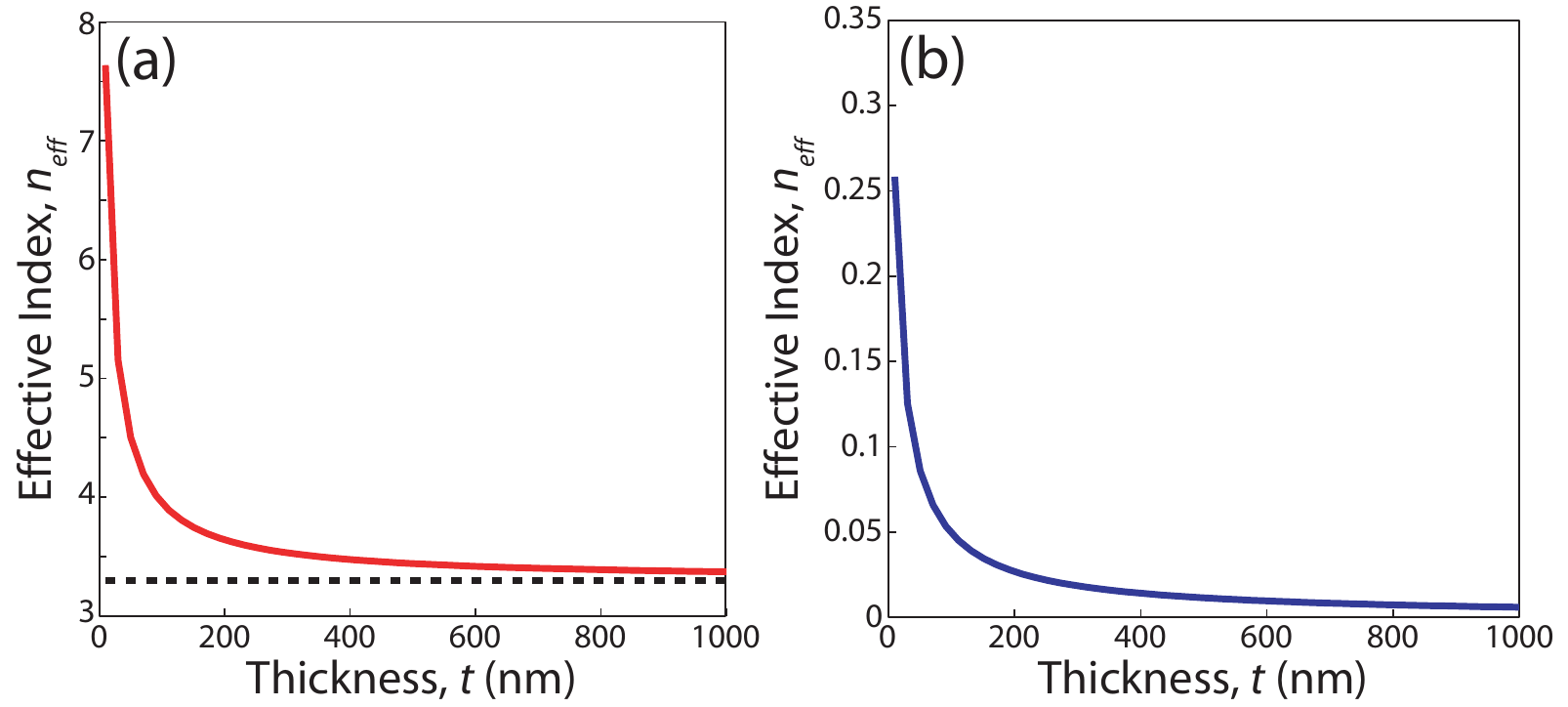}
\caption{The (a) real (red) and (b) imaginary (blue) parts of the dielectric constant dispersion relation of an Ag/GaAs/Ag waveguide vs. waveguide thickness $t$, for a free-space wavelength of $\lambda = 10~\mu$m. The GaAs core index is indicated by a dotted black line.} \label{fig:guidewidth_disp}
\end{center}
\end{figure}

In choosing the ideal waveguide thickness, $t$, for the double-metal (or metal-insulator-metal, MIM) plasmonic waveguide, there are several considerations. Figure \ref{fig:guidewidth_disp} shows the variation of mode effective index $n_{\textrm{eff}}$ with waveguide thickness for a Ag/GaAs/Ag waveguide at a free-space wavelength of $\lambda = 10~\mu$m. As the waveguide thickness decreases and more energy moves into the metal regions, both the real (Fig. \ref{fig:guidewidth_disp}(a)) and the imaginary (Fig. \ref{fig:guidewidth_disp}(b)) parts of the effective index increase. A high real part of the effective index is beneficial, because it increases the index contrast of the photonic crystal by increasing the contrast between the photonic crystal holes and the metal-covered regions (the double-metal waveguide). A higher index contrast increases the strength of the photonic crystal perturbation, improving the in-plane confinement $Q_{xy}$ of the resonator. If the index contrast is high enough, the photonic crystal holes can be etched only into the top metal, without removing material from the detector active region. In the double-metal case, this condition is even easier to achieve than in the single-metal case. However, the imaginary part of the effective index is proportional to the loss in the waveguide, and must be minimized. Therefore a waveguide width must be chosen to balance the competing factors of index contrast and loss.

\begin{figure}[h]
\begin{center}
\includegraphics[width=\columnwidth]{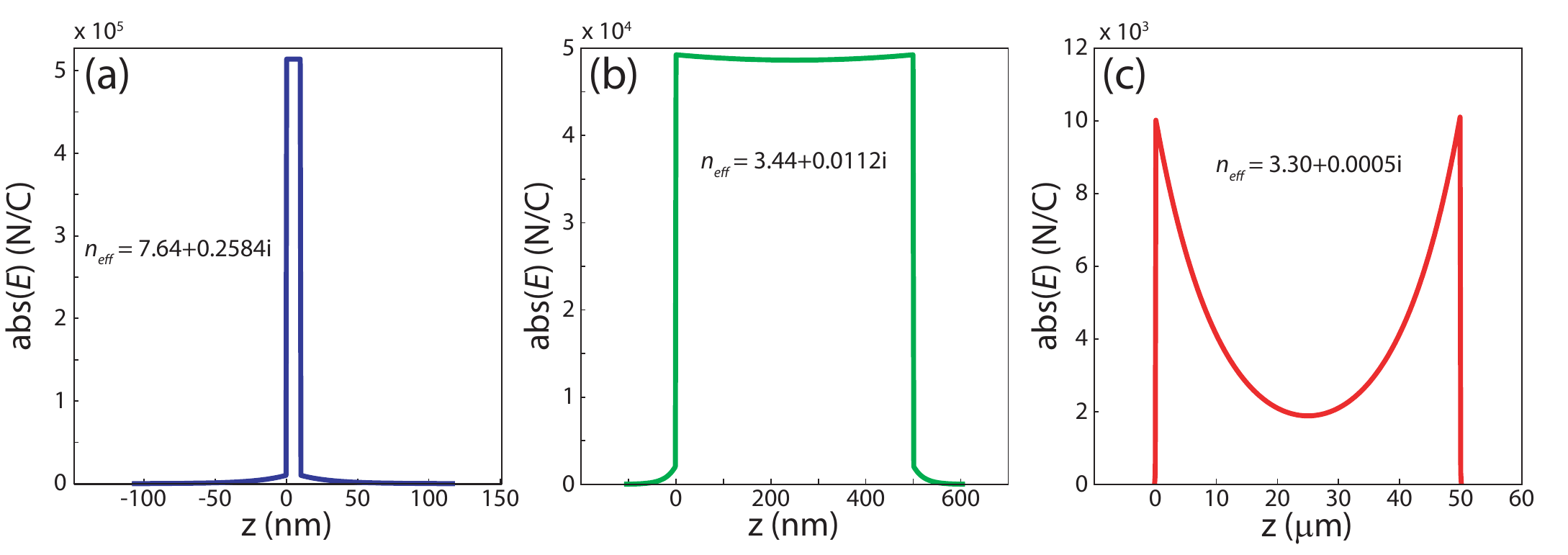}
\caption{The field profile for Ag/GaAs/Ag plasmon waveguides with a thickness $t$ of (a) 10 nm, (b) 500 nm, and (c) 50 $\mu$m are shown, for a free-space wavelength of $\lambda = 10~\mu$m. The effective index $n_{\textrm{eff}}$ for each plasmon waveguide is also given.} \label{fig:plasmon_profile}
\end{center}
\end{figure}

The field profiles of three Ag/GaAs/Ag plasmon waveguides are shown in Fig. \ref{fig:plasmon_profile}, with the calculated effective index $n_{\textrm{eff}}$, for a free-space wavelength of $\lambda = 10~\mu$m. Though the effective index values shown here would seem to generate only a low index contrast with the core dielectric material ($n_{GaAs} = 3.3$) for reasonable waveguide thicknesses, it must be considered that these simulations do not take into account the effect of the photonic crystal holes etched in the metal. In fact, it can be shown \cite{Pendry:2004, Bahriz:2007} that the presence of holes in a metal layer has the effect of lowering the effective plasmon frequency of that layer without significantly raising losses, increasing the real but not the imaginary part of the waveguide effective index. Thus, the actual combined photonic crystal and plasmon structure will have a considerably higher index contrast than would be expected from these effective index values. This is demonstrated via simulations of the full 3D structure which show that, even with the photonic crystal holes etched only into the top metal layer, we still achieve significant in-plane confinement and vertical coupling.

\begin{figure}[h]
\begin{center}
\includegraphics[width=\columnwidth]{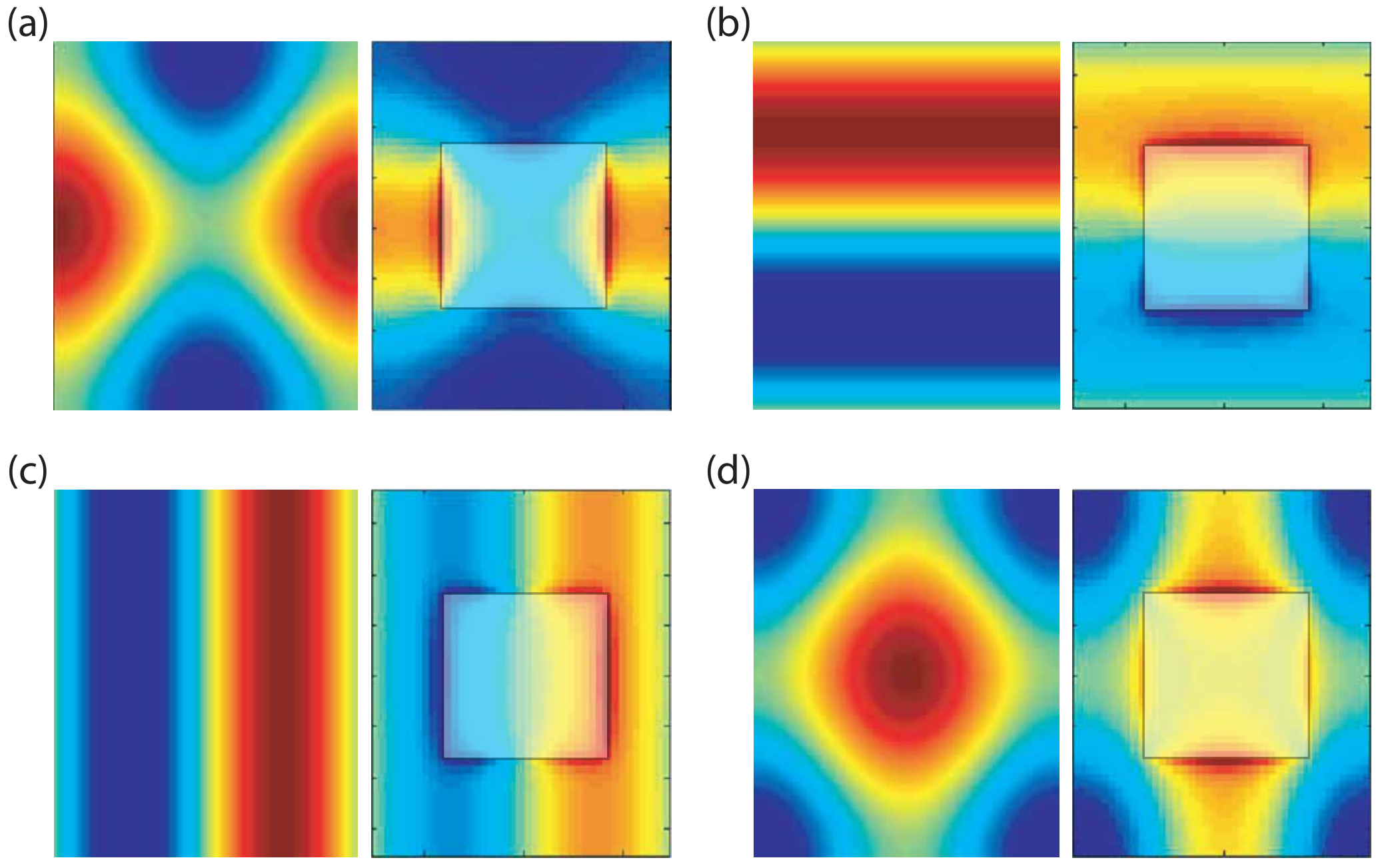}
\caption{A comparison of the FDTD simulated and group theory predicted $E_z$ field profiles for the four lowest gamma-point modes, in order of increasing frequency. The modes, labeled according to C$_{2v}$ designations, are (a) $A_{1,1}$, (b) $B_2$, (c) $B_1$, (d) $A_{1,2}$. The group theory predictions are shown on the left, while the FDTD results, including the effects of the photonic crystal air hole (overlaid white square), are shown on the right. The FDTD fields shown are 2D slices of the full simulations taken just below the top metal layer, inside the active region.} \label{fig:FDTD_group_theory_compare}
\end{center}
\end{figure}

Full structure simulations (photonic crystal plus plasmon waveguide) were performed using the FDTD method on a perfectly periodic lattice (as before, due to computational constraints). These FDTD simulations confirm the results of our separate photonic crystal and plasmon waveguide simulations. Figure \ref{fig:FDTD_group_theory_compare} shows the FDTD field plots (left) in comparison with the group theory mode plots (right). Though the presence of the air-hole distorts the shape of the modes in the center of the FDTD images, at the outside of the simulation region it can be seen that the simple group theory calculations have accurately predicted the mode shapes given by the more complex FDTD simulations.

\begin{figure}[h]
\begin{center}
\includegraphics[width=0.75 \columnwidth]{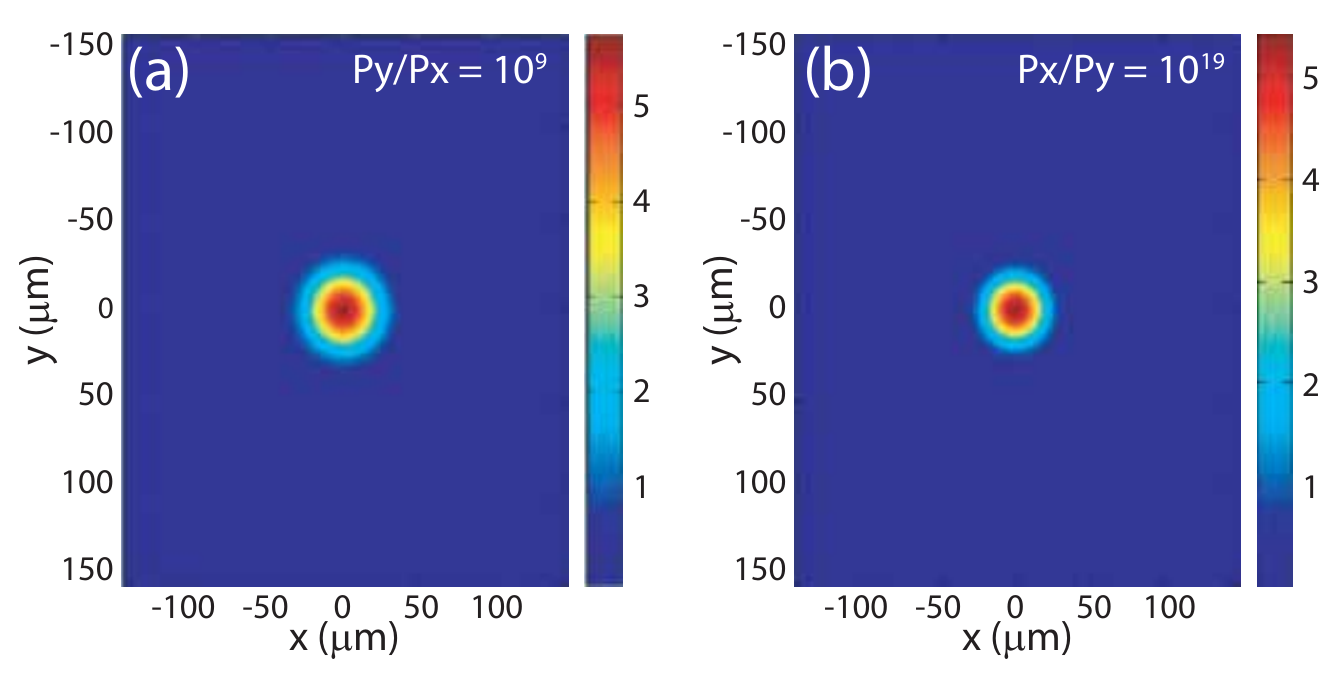}
\caption{Far-field plots at $z = 90~\mu$m of the two $\Gamma$-point dipole modes in a structure with $\bar{W} = 0.5309$. (a) $B_2$ mode power density, with dominant $\hat{y}$-polarization. (b) $B_1$ mode power density, with dominant $\hat{x}$-polarization. For both $B_1$ and $B_2$ modes, the polarization selectivity is calculated to be greater than $10^9$, limited entirely by error in the numerical simulation.} \label{fig:far_field}
\end{center}
\end{figure}

We have also investigated the far-field profiles of the four $\Gamma$-point modes, through examining the spatial fourier transform of $E_z$. Due to time-reversal symmetry, the field profile of a mode that can be coupled into the resonator is equivalent to the field profile of the resonator mode propagated out into the far-field, therefore these far-field plots indicate the mode-shapes and polarizations that couple most strongly from free space to the resonator mode. Far-field plots of the two fundamental stretched-lattice dipole modes, $B_1$ and $B_2$, are shown in Fig. \ref{fig:far_field}, generated from an 10X10 tiled array of the FDTD simulated field profile (itself one lattice constant in size) and apodized using a Gaussian function with a standard deviation of two lattice constants ($a_x$ and $a_y$, respectively) in the $x$ and $y$ directions. We can see from Fig. \ref{fig:far_field}(a,b) that the $B_1$ and $B_2$ modes are well-suited for coupling to incident free-space light, since the far-field profile has a single lobe at normal incidence and does not contain any anti-nodes, in agreement with the group theory predictions from Section~\ref{sec:photonic_crystal}.

\section{Critical Coupling}
\label{sec:critical_coupling}

\begin{figure}[h]
\begin{center}
\includegraphics[width=0.75 \columnwidth]{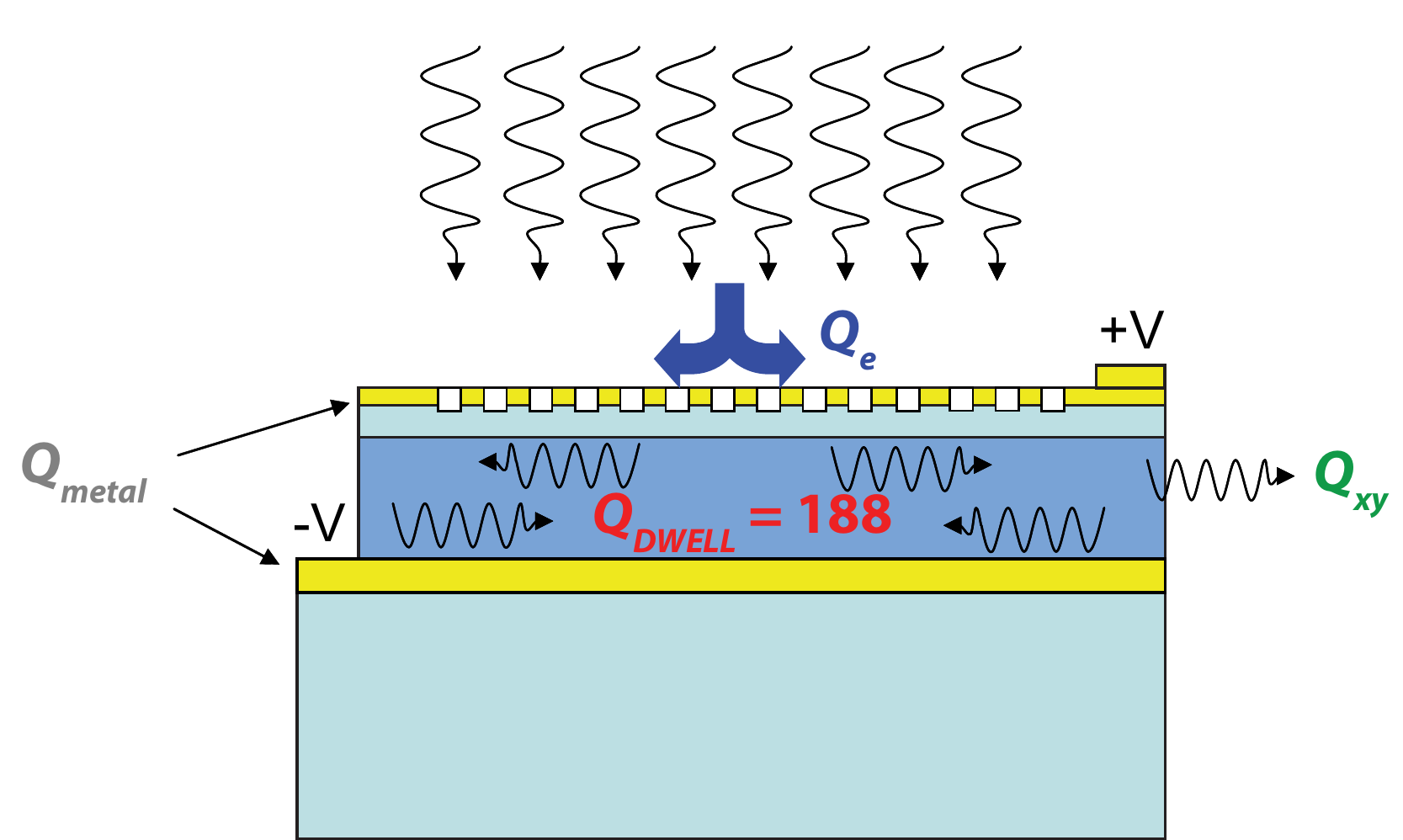}
\caption{The dominant loss mechanisms within a double-metal plasmonic photonic crystal resonant detector.} \label{fig:critical_coupling}
\end{center}
\end{figure}

After optimizing the large-scale resonator design and choosing photonic crystal modes which have the largest coupling to normal-incident light, it still remains to find the best values for the design parameters to increase detector absorption, and to determine the fundamental limits on absorption enhancement for these two (single-metal and double-metal) resonator designs. We find that, for the double-metal resonator, the absorption is greatest at the point of critical input coupling, whereas we find a different optimal point for the single-metal resonator, as the parasitic substrate loss increases along with the detector absorption as the input coupling is increased.

Critical coupling occurs when the external coupling to the resonator (the resonator ``coupling loss'') is equal to the total internal cavity loss from all other loss mechanisms. When that condition occurs, the reflection coefficient goes to zero, and all of the incident light at the resonance frequency is absorbed in the resonator \cite{Cai:2000}. Figure~\ref{fig:critical_coupling} shows the various scattering and absorption processes that are involved in near normal incidence resonant detection. The reflection from the cavity is given by
\begin{equation}
R = \frac{\Delta^2 + \left( \frac{\gamma_0 - \gamma_e}{2} \right)^2}{\Delta^2 + \left( \frac{\gamma_t}{2} \right)^2},
\label{eq:reflection}
\end{equation}
where $\Delta = \omega - \omega_0$ is the frequency detuning from the resonance frequency $\omega_0$, $\gamma_0$ is the intrinsic cavity loss rate, $\gamma_e$ is the vertical coupling rate to free space, and $\gamma_t = \gamma_0 + \gamma_e$. The loss rates $\gamma$ are related to the $Q$-factors given previously by $\gamma = n \lambda Q/2 \pi c$. It is clear from Eq.~\ref{eq:reflection} that, when the cavity is excited on-resonance ($\Delta = 0$), the reflection goes to zero when $\gamma_0 = \gamma_e$, when the rate at which the cavity can be fed from free space is equal to the sum of all the internal cavity loss rates. This is the critical coupling condition.

From the expression for the reflection in Eq.~\ref{eq:reflection}, we can write the power dropped into the cavity (not only the absorbed power, but all of the power not reflected):
\begin{equation}
P_d = P_{in} (1-R) = P_{in} \frac{\gamma_0 \gamma_e}{\Delta^2 + \left( \frac{\gamma_t}{2} \right)^2},
\end{equation}
where $P_{in}$ represents the power incident on the cavity. Therefore the fractional absorption efficiency into the $i$-th loss channel is
\begin{equation}
p_i = \frac{\gamma_i}{\gamma_0} \frac{P_d}{P_{in}} = \frac{\gamma_i \gamma_e}{\Delta^2 + \left( \frac{\gamma_t}{2} \right)^2}.
\end{equation}
We can enumerate the loss mechanisms in Fig.~\ref{fig:critical_coupling}, such that $\gamma_0 = \gamma_D + \gamma_{\textrm{metal}} + \gamma_{xy} + \gamma_{\textrm{sub}}$, corresponding to the (beneficial) detector absorption, the metal absorption, the in-plane loss, and the substrate loss, respectively. The in-plane loss can always be made negligible, by adding more lattice constants to the photonic crystal patterning region to increase the in-plane confinement strength relative to the other loss mechanisms. In the single-metal case, we can consider $\gamma_{\textrm{sub}} = m \gamma_e$, representing a mode coupling into the substrate that is $m$ times larger than that into the air due to the higher substrate refractive index; in the double-metal case, $m \sim 0$ due to the thick bottom layer of plasmon metal.

The fractional absorption into the DWELL detector material is then, at resonance, given by
\begin{equation}
p_{\textrm{D}} = \frac{4 \gamma_D \gamma{e}}{[(m+1) \gamma_e + \gamma_{\textrm{metal}} + \gamma_D]^2}.
\end{equation}
From this expression, we see that the maximum fractional absorption occurs at
\begin{equation}
\gamma_e = \frac{\gamma_D + \gamma_{\textrm{metal}}}{1+m}.
\end{equation}
As the input coupling $\gamma_e$ can be adjusted by varying resonator parameters, this maximal condition should be readily achievable, corresponding to a fractional absorption into the detector material of
\begin{equation}
p_{\textrm{D,max}} = \frac{\gamma_D}{(1+m)(\gamma_\textrm{D} + \gamma_{\textrm{metal}})}.
\end{equation}
For the double-metal case in which $m \sim 0$, the maximal detector absorption occurs at the pure critical coupling condition, $\gamma_e = \gamma_D + \gamma_{\textrm{metal}}$. In this case, the fractional power absorbed is primarily limited by the relatively small metal losses in the mid-infrared region. For the silver plasmon waveguides simulated in this work, we find a metal loss quality factor of $Q_{\textrm{metal}} = 149$, in comparison to the estimated DWELL detector absorption quality factor of $Q_D = 188$. This indicates that 55.8\% of the incoming light will be absorbed in the active material for the optimal external coupling quality factor of $Q_e = 83$.

\begin{figure}[h]
\begin{center}
\includegraphics[width=0.75 \columnwidth]{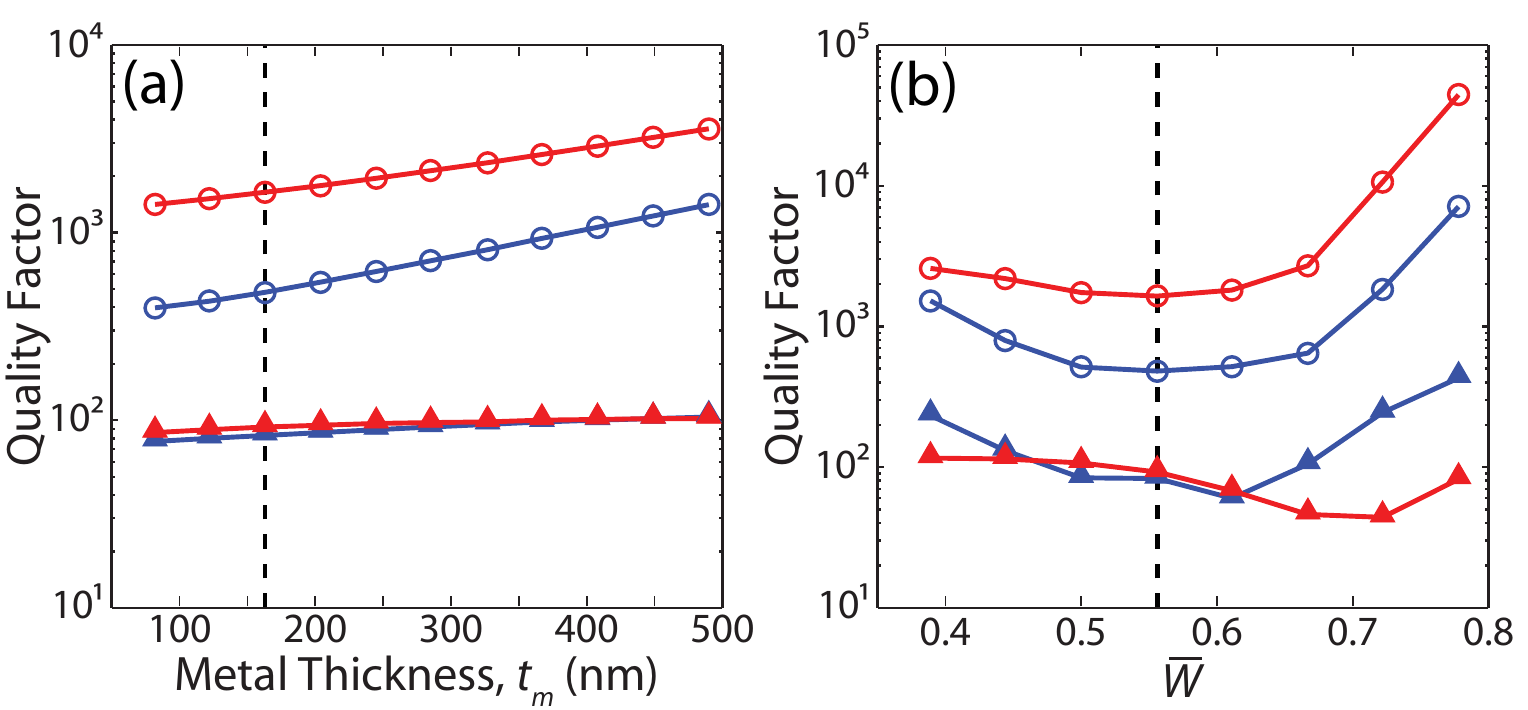}
\caption{(a) Variation of external coupling and substrate loss quality factors, $Q_e$ and $Q_{\textrm{sub}}$, with metal thickness $t_m$, and (b) with $\bar{W}$ for the fundamental (blue) and higher-order (red) modes of the unstretched single-metal photonic crystal lattice. Open circles represent $Q_e$ and filled triangles represent $Q_{\textrm{sub}}$. A dotted line marks the value of the parameter held constant in the opposing plot.} \label{fig:single_metal_thickness_width}
\end{center}
\end{figure}

\begin{figure}[h]
\begin{center}
\includegraphics[width=0.75 \columnwidth]{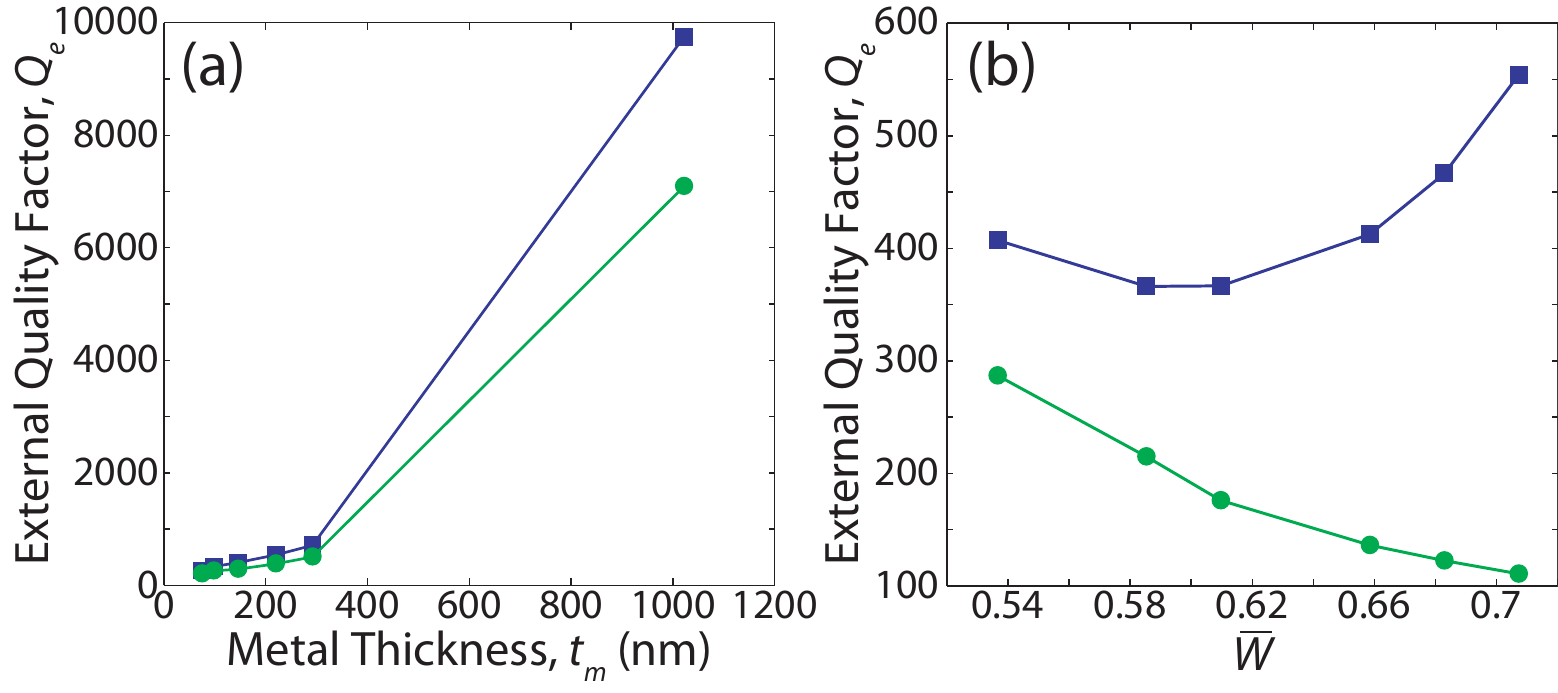}
\caption{Variation of vertical coupling quality factor $Q_e$ of the $B_1$ (blue square) and $B_2$ (green circle) modes of the double-metal photonic crystal lattice with changing (a) $t_m$ and (b) $\bar{W}$.} \label{fig:double_metal_thickness_width}
\end{center}
\end{figure}

There are many free parameters in this resonator structure which can be optimized in order to achieve the optimal input coupling value. Choosing two of the most significant, the normalized hole width, $\bar{W}$, and the top metal thickness, $t_m$, we investigate their effect on $Q_e$ for both the single-metal unstretched lattice (Fig.~\ref{fig:single_metal_thickness_width}) and double-metal stretched-lattice (Fig.~\ref{fig:double_metal_thickness_width}, with a lattice stretching ratio of 1.2) structures. With variation of $\bar{W}$, shown in Figs.~\ref{fig:single_metal_thickness_width}(b) for single-metal and \ref{fig:double_metal_thickness_width}(b) for double-metal, the overall trend for both structures is the same, showing a curve most likely due to a combination of factors: the increased hole size provides a larger aperture through which light can escape, decreasing the $Q_e$; but the larger air hole also distorts the shape of the mode, causing it to generate a less pure far-field profile which does not match as well with a free-space beam. In contrast, the variations in quality factor with changes in the top metal thickness, shown in Figs.~\ref{fig:single_metal_thickness_width}(a) for single-metal and \ref{fig:double_metal_thickness_width}(a) for double-metal, illustrate vertical quality factors $Q_e$ of both dipole-like modes increasing monotonically as the top metal becomes thicker. Varying the top metal thickness $t_m$ is an effective way to change $Q_e$ to better match the internal loss, and thus more closely approach critical coupling, without changing the mode frequency.

Though the behavior as hole size and metal thickness are varied is similar for both single-metal and double-metal structures, the double-metal structure has a lower achievable $Q_e$, indicating more favorable external coupling conditions; the stretching of the photonic crystal lattice does not significantly decrease $Q_e$.

\section{Conclusion}
\label{sec:conclusion} 

We have designed a plasmonic photonic crystal resonator, utilizing either a single-metal or double-metal plasmon waveguide, for use in mid-infrared photodetectors. This resonator design shows good frequency and polarization selectivity for use in hyperspectral and hyperpolarization detectors. We analyzed the conditions for optimal detector absorption enhancement, and by varying the photonic crystal hole size and top metal thickness, we adjusted the vertical coupling efficiency to more closely match the resonator loss, moving towards achieving critical coupling. Additional increases in coupling efficiency or reductions in loss could bring the system to near 100\% absorption in the detector. This resonator can be optimized for use at any wavelength from the terahertz to the visible with suitable scaling of the photonic crystal holes and waveguide width, and can easily be modified to suit any detector material, since no photonic crystal holes are etched into the active region itself.

\begin{figure}[h]
\begin{center}
\includegraphics[width=\columnwidth]{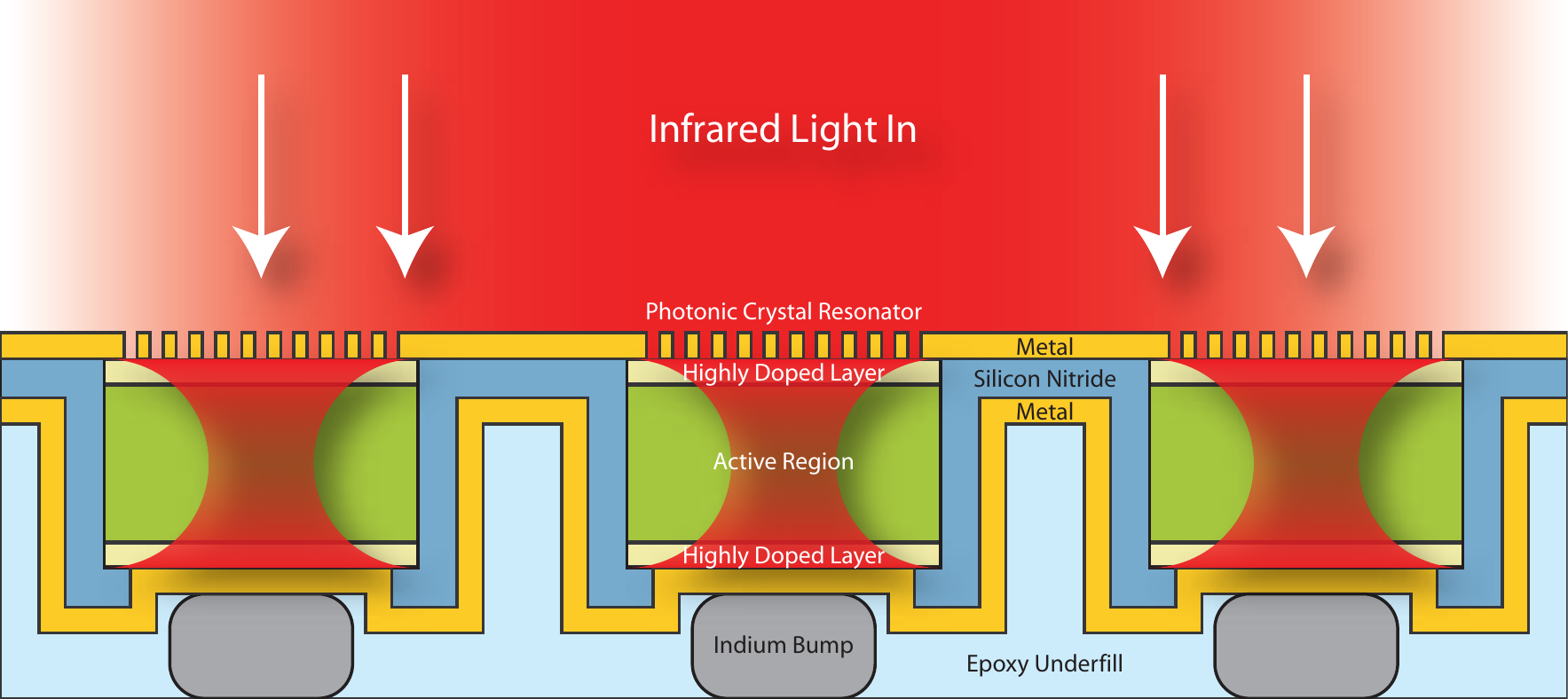}
\caption{A design schematic for a resonant double-metal plasmonic photonic crystal FPA.} \label{fig:fpa_schematic}
\end{center}
\end{figure}

The flip-chip bonding method of focal plane array (FPA) fabrication naturally lends itself to use with a double-metal resonant cavity, with only the top metal photonic crystal lithography step differing from standard process techniques. DWELL FPAs have already been demonstrated with hybridization to a readout integrated circuit \cite{Krishna:PhysD2005, Vandervelde:2008}. In Fig.~\ref{fig:fpa_schematic}, a proposed FPA schematic is shown, illustrating the ease with which double-metal plasmonic photonic crystal resonators can be incorporated into current FPA designs and presenting the possibility to achieve highly sensitive mid-infrared spectral and polarization imaging at low cost.

\section*{Acknowledgements}
\label{sec:Acknowledgements}

This work was supported by the AFOSR through grant \#FA9550-06-1-0443 and MURI grant \#FA9550-04-1-0434, the AFRL through grant \#FA9453-07-C-0171, and the IC post-doctoral program.

\end{document}